\documentclass[12pt]{article}
\usepackage{fancyhdr}

\usepackage{mathrsfs}
\usepackage[T1]{fontenc}
\usepackage{mathpazo}
\usepackage{setspace}
\usepackage{amsfonts}
\usepackage{amssymb}
\usepackage{amsmath}
\usepackage{epsfig}
\usepackage{latexsym}
\usepackage{color}
\usepackage{graphicx}
\usepackage{nicefrac}
\usepackage[latin1]{inputenc}
\usepackage{slashed}
\usepackage{multirow}
\usepackage{fancybox}
\usepackage{cite}
\usepackage{comment}
\usepackage{soul}

\usepackage{hyperref}

\def\hybrid{\topmargin -20pt    \oddsidemargin 0pt
        \headheight 0pt \headsep 0pt
        \textwidth 6.25in       
        \textheight 9.25in       
        \marginparwidth .875in
        \parskip 5pt plus 1pt   \jot = 1.5ex}

\hybrid

\def\baselinestretch{1.2}

\catcode`\@=11

\def\marginnote#1{}
\newcount\hour
\newcount\minute
\newtoks\amorpm
\hour=\time\divide\hour by60
\minute=\time{\multiply\hour by60 \global\advance\minute by-\hour}
\edef\standardtime{{\ifnum\hour<12 \global\amorpm={am}%
        \else\global\amorpm={pm}\advance\hour by-12 \fi
        \ifnum\hour=0 \hour=12 \fi
        \number\hour:\ifnum\minute<10 0\fi\number\minute\the\amorpm}}
\edef\militarytime{\number\hour:\ifnum\minute<10 0\fi\number\minute}

\def\draftlabel#1{{\@bsphack\if@filesw {\let\thepage\relax
   \xdef\@gtempa{\write\@auxout{\string
      \newlabel{#1}{{\@currentlabel}{\thepage}}}}}\@gtempa
   \if@nobreak \ifvmode\nobreak\fi\fi\fi\@esphack}
        \gdef\@eqnlabel{#1}}
\def\@eqnlabel{}
\def\@vacuum{}
\def\draftmarginnote#1{\marginpar{\raggedright\scriptsize\tt#1}}

\def\draft{\oddsidemargin -.5truein
        \def\@oddfoot{\sl preliminary draft \hfil
        \rm\thepage\hfil\sl\today\quad\militarytime}
        \let\@evenfoot\@oddfoot \overfullrule 3pt
        \let\label=\draftlabel
        \let\marginnote=\draftmarginnote
   \def\@eqnnum{(\theequation)\rlap{\kern\marginparsep\tt\@eqnlabel}%
\global\let\@eqnlabel\@vacuum}  }

\def\preprint{\twocolumn\sloppy\flushbottom\parindent 2em
        \leftmargini 2em\leftmarginv .5em\leftmarginvi .5em
        \oddsidemargin -.5in    \evensidemargin -.5in
        \columnsep .4in \footheight 0pt
        \textwidth 10.in        \topmargin  -.4in
        \headheight 12pt \topskip .4in
        \textheight 6.9in \footskip 0pt
        \def\@oddhead{\thepage\hfil\addtocounter{page}{1}\thepage}
        \let\@evenhead\@oddhead \def\@oddfoot{} \def\@evenfoot{} }

\def\numberbysection{\@addtoreset{equation}{section}
        \def\theequation{\thesection.\arabic{equation}}}

\def\underline#1{\relax\ifmmode\@@underline#1\else
        $\@@underline{\hbox{#1}}$\relax\fi}

\def\titlepage{\@restonecolfalse\if@twocolumn\@restonecoltrue\onecolumn
     \else \newpage \fi \thispagestyle{empty}\c@page\z@
        \def\thefootnote{\fnsymbol{footnote}} }

\def\endtitlepage{\if@restonecol\twocolumn \else \newpage \fi
        \def\thefootnote{\arabic{footnote}}
        \setcounter{footnote}{0}}  

\catcode`@=12
\relax

\def\figcap{\section*{Figure Captions\markboth
        {FIGURECAPTIONS}{FIGURECAPTIONS}}\list
        {Figure \arabic{enumi}:\hfill}{\settowidth\labelwidth{Figure
999:}
        \leftmargin\labelwidth
        \advance\leftmargin\labelsep\usecounter{enumi}}}
 \relax
\def\tablecap{\section*{Table Captions\markboth
        {TABLECAPTIONS}{TABLECAPTIONS}}\list
        {Table \arabic{enumi}:\hfill}{\settowidth\labelwidth{Table
999:}
        \leftmargin\labelwidth
        \advance\leftmargin\labelsep\usecounter{enumi}}}
 \relax
\def\reflist{\section*{References\markboth
        {REFLIST}{REFLIST}}\list
        {[\arabic{enumi}]\hfill}{\settowidth\labelwidth{[999]}
        \leftmargin\labelwidth
        \advance\leftmargin\labelsep\usecounter{enumi}}}
 \relax
\makeatletter
\newcounter{pubctr}
\def\publist{\@ifnextchar[{\@publist}{\@@publist}}
\def\@publist[#1]{\list
        {[\arabic{pubctr}]\hfill}{\settowidth\labelwidth{[999]}
        \leftmargin\labelwidth
        \advance\leftmargin\labelsep
        \@nmbrlisttrue\def\@listctr{pubctr}
        \setcounter{pubctr}{#1}\addtocounter{pubctr}{-1}}}
\def\@@publist{\list
        {[\arabic{pubctr}]\hfill}{\settowidth\labelwidth{[999]}
        \leftmargin\labelwidth
        \advance\leftmargin\labelsep
        \@nmbrlisttrue\def\@listctr{pubctr}}}
 \relax
\makeatother
\newskip\humongous \humongous=0pt plus 1000pt minus 1000pt

\newif\ifdtup

\relax

\def\be{\begin{equation}}
\def\ee{\end{equation}}
\def\ba{\begin{eqnarray}}
\def\ea{\end{eqnarray}}

\def\del{\partial}

\def\a{\alpha}

\def\b{\beta}

\def\g{\gamma}

\def\d{\delta}

\def\m{\mu}
\def\n{\nu}
\def\om{\omega}

\def\l{\lambda}
\def\L{\Lambda}
\def\s{\sigma}

\def\cN{{\cal N}}

\def\no{\noindent}

\def\qq{\qquad}

\def\IR{\relax{\rm I\kern-.18em R}}

\def \ha {{1\over 2}}

\def \ov {\over}

\def\diag{{\rm diag}}

\def\IR{\relax{\rm I\kern-.18em R}}
\def\IL{\relax{\rm I\kern-.18em L}}

\def\inv{^{\raise.15ex\hbox{${\scriptscriptstyle -}$}\kern-.05em 1}}

\def\Tr{{\rm Tr}}

\begin{document}

\renewcommand{\theequation}{\thesection.\arabic{equation}}
\csname @addtoreset\endcsname{equation}{section}

\newcommand{\eqn}[1]{(\ref{#1})}
\begin{titlepage}
\begin{center}

~
\vskip  .2in

{\bf \large Gauged WZW-type theories and the all-loop\\ anisotropic non-Abelian Thirring model}

\vskip 0.4in

{\bf Konstadinos Sfetsos}$^{1}$\phantom{x} and\phantom{x} {\bf Konstadinos Siampos}$^{2}$
\vskip 0.15 in
{\em ${}^1$Department of Nuclear and Particle Physics\\
Faculty of Physics, University of Athens\\
Athens 15784, Greece\\
{\tt ksfetsos@phys.uoa.gr}\\

\vskip 0.15in
${}^2$M\'ecanique et Gravitation, Universit\'e de Mons, 7000 Mons, Belgique\\
{\tt konstantinos.siampos@umons.ac.be}
}

\vskip .5in
\end{center}

\centerline{\bf Abstract}

\no
We study what we call the all-loop anisotropic bosonized Thirring $\s$-model. This
interpolates between the WZW model and the non-Abelian T-dual of the principal chiral model
for a simple group. It has an invariance involving the
inversion of the matrix parametrizing the coupling constants.
We compute the general renormalization group flow equations which
assume a remarkably simple form and derive its properties. For
symmetric couplings, they consistently truncate to previous results in the literature.
One of the examples we provide gives rise to a first order system of differential equations
interpolating between the Lagrange and the Darboux--Halphen integrable systems.

\newpage

\end{titlepage}
\vfill
\eject

\tableofcontents

\noindent

\vskip .4in

\def\baselinestretch{1.2}
\baselineskip 20 pt
\noindent


\setcounter{equation}{0}
\renewcommand{\theequation}{\thesection.\arabic{equation}}

\section{Introduction}
\label{intro}

The change in the behaviour of a field theoretical system is encoded in the way the
coupling constants of the theory alter with the energy scale. This is studied mathematically under the general frame
of the renormalization group (RG), a systematization started in the early seventies \cite{Wilson:1971bg}.
These investigations typically give rise to a system of
first order coupled non-linear differential equations, the RG flow or $\b$-function equations, for the couplings of the theory
(for a thorough introduction and a review of the subject see \cite{revrg}).
Typically one starts from an asymptotically free theory in the UV
or from a conformal field theory (CFT) and then flows away by perturbing with relevant operators.
In traditional approaches the RG flow equations are determined order by order in perturbation theory.
It is a natural question to ask if it is possible to compute these equations exactly in the coupling constants
or at least for some of them. This is important from a physics view point since one could discover new fixed point theories
towards the IR. Mathematically it is a very difficult task since in doing so one has to take into account irrelevant operators as
well.

Given the above comments it is always exciting if we can obtain the exact RG flow equations for (at least some of) the couplings of a theory.
Intuitively we expect that this could be feasible if the perturbed theory is highly symmetric. Such cases arise when the starting point is a
two-dimensional CFT with infinitely dimensional current algebra symmetries for the left and the right movers.

A model where this is possible to a certain extend is the bosonized version \cite{Witten:1983ar,Karabali:1988sz} (and references therein)
of the non-Abelian Thirring model \cite{Dashen:1974gu}. We will call in short this the
non-Abelian Thirring model which has an action of the form
\be
S = S_0 + {k \l\ov \pi} \int J^a_+ J^a_-\ ,
\label{fljh2s}
\ee
where $S_0$ describes a CFT containing right and left affine Lie algebras both at level $k$
with currents $J^a_+$ and $J^a_-$, respectively.
The $\b$-function for this theory was computed in \cite{Kutasov:1989dt} to all orders in $\l$ and to leading order in $1\ov k$.
The generalization of this computation to the anisotropic non-Abelian Thirring model in which the current-current interaction in \eqn{fljh2s} is replaced by
an arbitrary symmetric coupling matrix was performed in \cite{Gerganov:2000mt}.
In these cases the computations were performed using current-algebra techniques without
much reference to the geometrical details of the background in \eqn{fljh2s}. It is very interesting
to obtain an effective action in which all effects of the parameter $\l$ have been incorporated exactly and where
 the only perturbative expansion is with respect to $1/k$.

In a recent development a large family of $\s$-models was constructed in \cite{Sfetsos:2013wia} by a gauging procedure.
It interpolates between the Wess--Zumino--Witten (WZW) model and
the non-Abelian T-dual of the principal chiral model (PCM) model for a simple group $G$.
In the simplest case this $\s$-model action was shown to be
integrable \cite{Sfetsos:2013wia} by demonstrating that certain algebraic constraints for integrability \cite{Balog:1993es} were satisfied.
It incorporates non-trivially a single parameter, for small values of which it coincides with \eqn{fljh2s}.
Remarkably, the RG flow equation to leading order in the $1/k$ expansion was computed in \cite{Itsios:2014lca}
and coincides with the one in \cite{Kutasov:1989dt}. Based on that, it was proposed that this action is the all loop
effective of the non-Abelian Thirring model \eqn{fljh2s}. In further support, both actions
share the same global symmetries and roughly speaking both possess an additional symmetry under the inversion of the deformation parameter $\l$.
This is manifest for the action of \cite{Sfetsos:2013wia} but also arises implicitly from path integral considerations
involving \eqn{fljh2s} and symmetry arguments, in \cite{Kutasov:1989aw}.

The initial aim of this work was to compute the RG flows for the anisotropic $G=SU(2)$ case, with diagonal coupling matrix $\l_{ab}$
and then to compare the results with the analogue ones in \cite{LeClair:2004ps} which were found using \cite{Gerganov:2000mt}.
Nevertheless, we managed to compute the RG flow equations for the most general class of the $\s$-models
of \cite{Sfetsos:2013wia} containing a general deformation matrix $\l_{ab}$ and for general simple group $G$,
and present the result in a remarkably simple and compact form.
In the above $SU(2)$ case the derived RG-flow interpolate, between the Lagrange and Darboux--Halphen integrable systems,
naturally explained by the interpolating nature of our $\s$-models.

We believe, but we do not have a proof, that our general RG-flow equations coincide with those of \cite{Gerganov:2000mt,LeClair:2000ya} when such a
comparison can be made, i.e. when $\l_{ab} = \l_{ba}$. However, we provide three non-trivial examples, and found agreement with results following by using the expressions of \cite{Gerganov:2000mt,LeClair:2000ya}. In that sense the general $\s$-model of \cite{Sfetsos:2013wia} can be thought of as the effective action for the most
general anisotropic non-Abelian Thirring model where all effects related to the coupling matrix $\l_{ab}$ (which may have
an antisymmetric part) have been taken into account.
As a byproduct of our work we obtain the RG flow equations of the non-Abelian T-dual of the general PCM and we prove that they
match those of the general PCM.

The organization of this paper is as follows: In section \ref{frame} we set up the general class of interpolating $\s$-models
as suited for our purposes. In section
\ref{Symmetries} we study various of its symmetries and limits as well as the resulting constraints for the RG flow equations
for the coupling matrix $\l_{ab}$.
In section \ref{geometry} we compute the generalized spin-connection and Ricci tensor corresponding to the metric and
antisymmetric tensor of our $\s$-models. In section
\ref{Tour} we derive and study the $\b$-function for the couplings $\l_{ab}$.  In section \ref{Applications}
we present two examples based on the anisotropic $SU(2)$ and on the symmetric coset $G/H$ space.
In section \ref{Comparison} we compare with existing literature results.
We end up our work with a wrap up and a discussion on future directions in section \ref{conclusion}.

\setcounter{equation}{0}
\renewcommand{\theequation}{\thesection.\arabic{equation}}

\section{Setting the frame}
\label{frame}

In this section we present the two-dimensional $\s$-models of interest to us, in a way
suitable for studying their behaviour under RG flow in subsequent sections.

We will study the $\s$-models of \cite{Sfetsos:2013wia} which we first briefly review for the reader's convenience.
Consider a general compact simple group $G$ and a corresponding group element $g$ parametrized by $X^\m$, $\m=1,2,\dots , \dim(G)$.
The right and left invariant Maurer--Cartan forms, as well as the orthogonal matrix relating them, are defined as
\be
\begin{split}
& J^a_+ = -i\, {\rm Tr}(t^a \del_+ g g^{-1}) = R^a_\m \del_+ X^\m \ ,\qq J^a_- = -i\, {\rm Tr}(t^a g^{-1} \del_- g )= L^a_\m \del_- X^\m\ ,\\
& R^a = D_{ab}L^b \ ,\qq D_{ab}={\rm Tr}(t_a g t_b g^{-1})\ ,
\label{jjd}
\end{split}
\ee
which obey
\be
\mathrm{d}L^a=\frac12\,f_{abc}\,L^b\wedge L^c\,,\qq \mathrm{d}R^a=-\frac12\,f_{abc}\,R^b\wedge R^c\,,\qq
\mathrm{d}D_{ab}=D_{ac}\,f_{cbe}\,L^e\,.
\ee
The matrices $t^a$ obey the commutation relations $[t_a,t_b]=i f_{abc} t_c$
and are normalized as ${\rm Tr}(t_a t_b)=\d_{ab}$.
Then the form of the general $\s$-model action is given by \cite{Sfetsos:2013wia}
\be
S_{k,E}(g) = S_{{\rm WZW},k}(g) + {{k^2}\ov \pi}  \int J_+^a (E-k(D^T-\mathbb{I}))^{-1}_{ab}J_-^b\ ,
\label{skE}
\ee
where $E$ is a real matrix parametrizing the coupling constants of the theory. The first term is
the WZW action for a group $G$ which can be explicitly written as\footnote{
The relative coefficient of the quadratic and cubic terms is completely dictated by the Polyakov--Wiegmann (PW) formula
\cite{Polyakov:1984et}
\begin{equation*}
S_{{\rm WZW},k}(g_1 g_2) = S_{{\rm WZW},k}(g_1) + S_{{\rm WZW},k}(g_2)
- {k\ov \pi} \int \Tr(g_1^{-1} \del_- g_1 \del_+ g_2 g_2^{-1})\ ,
\end{equation*}
which is also very practical in evaluating the action for specific
parametrizations of $g\in G$.}
\be
S_{{\rm WZW},k}(g) = \frac{k}{2\pi}
\int L^a_\mu L^a_\nu\, \del_+X^\mu\del_-X^\nu+\frac{k}{12\pi}\int_B f_{abc} L^a\wedge L^b\wedge L^c\ .
\ee
The $J^a_\pm$ are the chirally and antichirally conserved currents of the WZW model.

\no
It is better for our purposes to reparametrize the couplings by introducing
the matrix
\be
\l= k (k \ \mathbb{I} + E)^{-1}\ .
\label{defl}
\ee
Then the action \eqn{skE} becomes
\be
S_{k,\l}(g) = S_{{\rm WZW},k}(g) + {{k}\ov \pi}  \int J_+^a (\l^{-1}-D^T)^{-1}_{ab}J_-^b  \ .
\label{tdulalmorev2}
\ee
If the matrix $\l$ is proportional to the identity, then the corresponding $\s$-model is
of special interest since it is actually integrable.
This was proven in \cite{Sfetsos:2013wia} by showing that the corresponding metric and antisymmetric tensor fields satisfy the
algebraic constraints for integrability of \cite{Balog:1993es} and \cite{Evans:1994hi}.
In addition, in \cite{Evans:1994hi} an $S$-matrix for the $SU(2)$ case was put forward and checked successfully against perturbation theory.
A form of the action similar to \eqn{tdulalmorev2} has appeared before in \cite{Tseytlin:1993hm}, along with related to this action discussion.

\section{Symmetries and limits of the RG flow}
\label{Symmetries}
In this section we study various symmetries, properties and limits of \eqref{tdulalmorev2} as well as
the emerging constraints on the RG flow.

The action \eqn{tdulalmorev2} has a remarkable symmetry under the inversion of the matrix $\l$, of the group
element $g$ and a simultaneous flip of the sign of the overall scaling $k$. This is encoded mathematically in
the relation
\be
\label{symmetry.sigma}
S_{-k,\l^{-1}}(g^{-1}) = S_{k,\l}(g)\ .
\ee
We note that both terms in \eqn{tdulalmorev2}
with the precise coefficients are necessary for the proof, which is
otherwise quite straightforward.

\no
Consider the limit where all the entries of the coupling matrix $\l$ are small and go to zero at the
same rate, i.e. the ratio of any two entries is finite. In this limit the action \eqn{tdulalmorev2}
can be approximated by
\be
S_{k,\l}(g) = S_{{\rm WZW},k}(g) + {k\ov \pi} \int \l_{ab} J_+^aJ_-^b +{\cal O}(\l^2)\ ,
\label{thorio}
\ee
corresponding to the WZW theory perturbed by the current bilinear $J_+^aJ_-^b$ with arbitrary coupling matrix $\l_{ab}$.
The first two terms define what we will call the anisotropic non-Abelian Thirring model in analogy
with the non-Abelian Thirring model \cite{Dashen:1974gu}, \cite{Karabali:1988sz}.
It becomes already apparent in this limit that the left-right current algebra symmetry of the WZW model breaks down
completely for a generic matrix $\l$ implying that the $\s$-model \eqn{tdulalmorev2} has no isometries whatsoever.
However, if we allow for a simultaneous transformation of the coupling matrix $\l$ then the
$\s$-model action \eqn{tdulalmorev2} (and of course its limit \eqn{thorio}) is invariant under
\be
g\to g_0^{-1} g g_0 \ ,\qq \l \to D_0^T \l D_0\ ,
\label{gg0g}
\ee
where $g_0\in G$ is a constant group element and $D_0$ is defined in \eqn{jjd} using $g_0$.
If $\l$ is invariant under the above transformation for all constant group elements $g_0$
then, from Schur's lemma, $\l$ is necessarily proportional to the unit matrix, i.e. $\l_{ab} = \l \d_{ab}$.
In that case the action \eqn{tdulalmorev2} becomes
\be
S_{k,\l}(g) = S_{{\rm WZW},k}(g) + {k\l \ov \pi}  \int J_+^a (\mathbb{I}-\l D^T)^{-1}_{ab}J_-^b  \
\label{tdulalmorev2orio}
\ee
and has a true isometry associated with the transformation $g\to g_0^{-1} g g_0$.

We believe that this action can be uniquely determined under certain assumptions and by
symmetry considerations. The argument goes as follows:
assuming that such an action contains the WZW action term and that any additional term
preserves two-dimensional Lorentz invariance and contains two world-sheet derivatives,
implies a term of the form $F(D)_{ab} J^a_+ J^b_-$.
The matrix $F(D)$ has to transform covariantly under the global symmetry $g\to g_0^{-1} g g_0$
(perhaps accompanied by a transformation of the coupling constants that $F(D)$ contains).
If we subsequently demand that $F(D)$ contains a single coupling parameter $\l$ and invariance
under the symmetry \eqn{symmetry.sigma} then we find no other possibility but the action \eqn{tdulalmorev2orio}.
Generalizing to a general coupling matrix $\l$
leads rather straightforwardly to the more general action \eqn{tdulalmorev2orio}.

\no
The symmetry \eqn{symmetry.sigma} is very powerful and restricts the form of the RG flow equations
for the $\l_{ab}$'s. The corresponding $\b$-function at one-loop
in the $1/k$ expansion is clearly of the form
\be
\b_\l ={\mathrm{d}\l\ov \mathrm{d}t} = - {f(\l)\ov k}\ ,
\label{dhdfl}
\ee
where $ t = \ln \m$, with $\m$ being the energy scale and where $f(\l)$ is a matrix to be determined
as we will explicitly do so in section 4. Here we note that
due to the symmetry \eqn{symmetry.sigma} we have the relation
\be
\label{symmetry.beta}
\l f(\l^{-1})\l = f(\l) \ ,
\ee
which severely constrains the matrix $f(\l)$. \\
In fact when $\l_{ab}=\l \d_{ab}$ this symmetry together
with CFT arguments allowed for the almost complete determination of $\b_\l$.
In this case, the $\b$-function computed in \cite{Itsios:2014lca}
coincides with the all-loop (and leading order in $1/k$) result for the non-Abelian Thirring model
found in \cite{Kutasov:1989dt}.
The map $\l\to1/\l$ and $k\to-k$ (for large values of $k$) was also noted in \cite{Kutasov:1989aw} without however an explicit
realization for the action such as the one in \eqn{tdulalmorev2orio}. It was rather deduced though path integral
and symmetry considerations.

\no
In addition, invariance under the symmetry \eqn{gg0g} implies that
\be
\displaystyle f(D_0^T \l D_0) = D_0^Tf(\l)D_0 \ .
\ee

Finally, we note that for $k\gg 1$ we may show that the action \eqn{tdulalmorev2} becomes
the non-Abelian T-dual of the $\s$-model action
\be
S_{\rm PCM} = {1\ov \pi} \int E_{ab} L^a_+ L^b_- = {1\ov \pi} \int E_{ab} L^a_\m L^b_\n\, \del_+ X^\m \del_- X^\n \ ,
\label{spcm}
\ee
which is the PCM action with general coupling matrix $E_{ab}$, a fact proven in \cite{Sfetsos:2013wia}.
We reproduce the proof here for the reader's convenient.
Expanding the matrix elements of $\l$ in \eqn{defl} near the identity we have that
\be
\l_{ab} =\d_{ab} - {1\ov k} E_{ab} + {\cal O}\left(1\ov k^2\right)\ .
\label{laborio}
\ee
To get a finite result in the limit $k\to \infty$ in the action \eqn{tdulalmorev2}, one is forced
to also expand the group element near the identity as
\be
g = \mathbb{I} + i {v\ov k} + {\cal O}\left( 1\ov k^2 \right)\ , \qq v = v_a t^a\ .
\label{nonbba}
\ee
This effectively introduces a non-compact set of variables $v_a$ in place of the original ones $X^\m$.
In that limit we have that
\be
J_\pm^a = {\del_\pm v\ov k} + {\cal O}\left( 1\ov k^2 \right)  \ ,\qq
D_{ab} = \d_{ab} + {f_{ab} \ov k} + {\cal O}\left( 1\ov k^2 \right) \ ,\qq f_{ab} = f_{abc} v_c\ .
\ee
Then in the limit $k\to \infty$ the action \eqn{tdulalmorev2} becomes
\be
S_{\rm non\!-\!Abel}(v) = {1\ov \pi} \int \del_+ v^a (E + f)^{-1}_{ab}\del_- v^b  \ ,
\label{nobag}
\ee
which is indeed the non-Abelian T-dual of the PCM action \eqn{spcm} \cite{Sfetsos:1996pm} (for the case with
$E_{ab}=\d_{ab}$ this was shown before for $SU(2)$ in \cite{Curtright:1994be} and for a general group
in \cite{Lozano:1995jx}).

\section{The generalized spin connection and Ricci tensor}
\label{geometry}

In this section we compute the generalized spin connection that includes the torsion
and the associated Ricci tensor for the background corresponding to the $\s$-model \eqn{tdulalmorev2}. These
are necessary in order to determine the $\beta$-function equations for the couplings $\l_{ab}$.
Our computation will parallel, in a sense, the one in \cite{Itsios:2014lca} for the case with $\l_{ab}=\l \d_{ab}$.

From the metric corresponding to the action \eqn{tdulalmorev2} we extract the frame fields
\begin{equation}
\label{frames}
  \mathrm{d}s^2=\begin{cases}
    g_{ab}e^ae^b\,,\qq g_{ab}=(\mathbb{I}-\l^T\l)_{ab}\,,\qq e^a=(D-\l)^{-1}_{ab}R^b\,, & \\
    \tilde g_{ab}\tilde e^a\tilde e^b\,,\qq \tilde g_{ab}=(\mathbb{I}-\l\l^T)_{ab}\,,\qq \tilde e^a=(D^T-\l^T)^{-1}_{ab}L^b\,. &
  \end{cases}
\end{equation}

Hence, depending on the frame we will use, we will bear in mind the use of the metric in raising and lowering indices.
This will be done with the metrics $g_{ab}$ and $\tilde g_{ab}$ and their inverses which will
be denoted by $g^{ab}=g^{-1}_{ab}$ and $\tilde g^{ab}=\tilde g^{-1}_{ab}$.
It turns out that the corresponding background is non-singular if and only if these metrics are positive-definite.
We also note that the frame fields transform under \eqref{symmetry.sigma} as $(e,\tilde e)\mapsto (\l e,\l^T \tilde e)$ and that the
metric picks up an overall minus sign since we have not included $\tfrac{k}{2\pi}$ in its definition.

\no
The matrix relating the above frames reads
\be
\label{Lambda}
\tilde e^a=\L_{ab} e^b\,,\qq \L=(\mathbb{I}-D\l^T)^{-1}(D-\l)=(D^T-\l^T)^{-1}(\mathbb{I}-D^T\l)\ .
\ee
It transforms under \eqref{symmetry.sigma} as $\L\mapsto\l^T\L\l^{-1}$  and satisfies the condition\footnote{
To prove this we found useful to make use of the identity
$$\mathbb{I}-\l\l^T=(\mathbb{I}-\l D^T)(\mathbb{I}-D\l^T)+(D-\l)\l^T+\l(D^T-\l^T)\,.$$
}
\be
\label{ortho}
\L^T(\mathbb{I}-\l\l^T)\L=\mathbb{I}-\l^T\l\ .
\ee
We note that $\L$ is an orthogonal matrix only when $\l$ is proportional to the identity.

From the action \eqn{tdulalmorev2} we also read off the antisymmetric two-form
\be
\label{2form}
B=\frac{k}{\pi}\left(B_0+R^T\l\wedge e\right)=\frac{k}{\pi}\left(B_0-L^T\l^T\wedge \tilde e\right)\,,
\ee
where $B_0$ is the antisymmetric two-tensor corresponding to the WZW model action.
The three-form field strength associated to $B_0$ is
\be
H_0=-\frac16\, f_{abc} L^a\wedge L^b\wedge L^c=-\frac16\, f_{abc} R^a\wedge R^b\wedge R^c\ .
\ee

\no
In the following we will use the $\tilde e^a$ frame and we will notationally
supply the geometric quantities associated with it with a tilde.
The spin-connection $\widetilde\omega_{ab}$
is equal to
\be
\label{spin}
\begin{split}
&\widetilde\omega_{ab}=\widetilde\omega_{ab|c}\tilde e^c\,,\qq \widetilde\omega_{ab}=-\widetilde\omega_{ba}\,,\quad
\widetilde\omega_{ab|c}=\frac12\left(\widetilde C_{abc}-\widetilde C_{cab}+\widetilde C_{bca}\right)\,,\\
&\mathrm{d}\tilde e^a=\frac12\,\widetilde C^a{}_{bc}\tilde e^b\wedge\tilde e^c\,,\quad
\widetilde C^a{}_{bc}:=-\widetilde C^a{}_{cb}\,,\quad \widetilde C_{abc}:=\tilde g_{ad}\,\widetilde C^d{}_{bc}\,.
\end{split}
\ee
A simple computation shows that
\be
\mathrm{d}\tilde e^a=-\frac12(D^T-\l^T)^{-1}_{ab}f_{bcd}(D^T-\l^T)_{ce}(D^T+\l^T)_{df}\,\tilde e^e\wedge \tilde e^f\ .
\ee
Using $f_{abc}D_{ia}D_{jb}D_{kc}=f_{ijk}$ and the first of the identities
\be
(D^T-\l^T)^{-1}=(1-\l\l^T)^{-1}(\L^{-T}+\l)\ ,\quad (D^T-\l^T)^{-1}=(\L+\l)(\mathbb{I}-\l^T\l)^{-1}\ ,
\ee
with notation $\L^{-T}=(\L^{-1})^T$, we may further write that
\be
\begin{split}
\label{dtilde}
& \mathrm{d}\tilde e^a=-\frac12\,f_{abc}\tilde e^b\wedge \tilde e^c-\frac12\left((1-\l\l^T)^{-1}(\L^{-T}+\l)\right)_{ad}\times\\
& \phantom{xxxxxx} \left(\l_{ed}\,f_{ebc}-\l_{be}\l_{cf}f_{def}\right)\tilde e^b\wedge \tilde e^c\ .
\end{split}
\ee
After some further manipulations we arrive at
\be
\label{dtildee}
\mathrm{d}\tilde e^a=-\frac12\,\tilde g^{am}\left(f_{mbc}-\l_{me}\l_{bn}\l_{c\ell}\,f_{en\ell}+
\L^{-T}_{mf}\left(\l_{ef}f_{ebc}-\l_{bn}\l_{c\ell}\,f_{fn\ell}\right)\right)\tilde e^b\wedge \tilde e^c\ ,
\ee
from which we compute $\widetilde \om_{ab|c}$.

Next we turn to the computation of the field strength of the two-form \eqref{2form}.
Using the identities
\be
\begin{split}
& (\mathbb{I}-\l\l^T)(\L+\l)(\mathbb{I}-\l^T\l)^{-1}=\L^{-T}+\l\,,\\
&  (D-\l)\l^T\L(D-\l)^{-1}=\L^{-T}\l^T\ ,
\end{split}
\ee
this is found to be
\be
\label{dB}
H=-\frac16\Big[f_{abc}-\l_{ad}\l_{be}\l_{cf}\,f_{def}+3\L^{-T}_{cf}\left(\l_{mf}f_{abm}
-\,\l_{ad}\l_{be}\,f_{def}\right)\Big]\,\tilde e^a\wedge \tilde e^b\wedge \tilde e^c\ .
\ee
Then the generalized spin connection $\widetilde \omega^-_{ab}$ that includes the torsion
\be
\widetilde\omega^-_{ab}=\widetilde\omega^-_{ab|c}\,\tilde e^c\,,\qq \widetilde\omega_{ab|c}^-=\widetilde\omega_{ab|c}-\frac12\widetilde H_{abc}\ ,
\ee
is found to be 
\be
\label{spinminus}
\widetilde\omega_{ab|c}^-=\L^{-T}_{cd}\left(\l_{md}f_{mab}-\l_{am}\l_{bn}f_{dmn}\right)\ .
\ee
To proceed with the computation of the Ricci tensor we require the exterior derivative of the matrix $\L^{-T}$.
We found that
\be
\begin{split}
\label{dLambda}
&\mathrm{d}\L^{-T}_{ab}=(\mathbb{I}-\l^T\l)^{-1}_{mb}\left(\left(\l_{dm}f_{adf}-\l_{ac}\l_{fe}f_{cme}\right)-
\L_{dm}\left(f_{daf}-\l_{dn}\l_{ac}\l_{fe}f_{nce}\right)\right.\\
&\left. \phantom{xxxxx}
+\L_{ac}^{-T}\left(\l_{fe}f_{mce}-\l_{dm}\l_{nc}f_{dnf}\right)-\L^{-T}_{ac}\L_{dm}\left(\l_{nc}f_{dnf}-\l_{dn}\l_{fe}f_{nce}\right)\right)\tilde e^f\,.
\end{split}
\ee
Finally we compute the generalized Ricci tensor by employing the general formula
for an antisymmetric spin-connection
\be
\label{Ricci.identity}
R^{\pm}_{ab}=\partial_c\,\om^{\pm c}{}_{a|b}-
\om^{\pm}_{ac|d}\,\om^{\mp}_b{}^{d|c}-\nabla^{\pm}_b\om^{\pm c}{}_{a|c}\,,\quad
\om_{ab|c}^{\pm}-\om_{ac|b}^{\mp}= C_{abc}\,,
\ee
where $\partial_a= e_a{}^\mu\partial_\mu.$
We have used this particular form since, as will see, it will make manifest the appearance of diffeomorphism terms
in the RG flow of the coupling matrix elements $\l_{ab}$.
Using \eqn{Ricci.identity} and \eqref{dtildee}, \eqref{spinminus} and \eqn{dLambda} after some manipulations we find that
the generalized Ricci tensor is
\be
\label{Ricci.general}
\widetilde R^-_{ab}=-\left(\l_{\ell i}f_{a\ell p}-\l_{aq}\l_{p\ell}f_{qi\ell}\right)\left(\l_{ce}f_{rme}-\l_{nr}\l_{dm}f_{ndc}\right)g^{im}\tilde g^{pc}\,\L^{-1}_{rb}
-\widetilde \nabla^-_b\widetilde\om^{-c}{}_{a|c}\ .
\ee

Note that in computing the relevant part of the generalized Ricci tensor for the RG flow
it was not necessary to know the precise form of
$\widetilde R^-_{ab}$, let alone to know first the exact expression of the generalized Riemann tensor.
These would have required
a much more involved computation. The reader will appreciate this remark
if he or she gives a glance at  eqs. (A.11) and (A.12) of \cite{Itsios:2014lca} for the simplest case with $\l_{ab}=\l\d_{ab}$.

\subsection{Single coupling}

We specialize to the case with  $\l_{ab}=\l\,\d_{ab}$.
Then the previous expressions simplify drastically and we find that
\begin{equation}
\begin{split}
&\mathrm{d}\tilde e^a=-\frac12\,\left(c_1+c_2\L\right)_{ab}f_{bcd}\tilde e^c\wedge \tilde e^d\,,\\
&H=-(1-\l^2)\left(\frac{c_1}{6}\,f_{abc}\,\tilde e^a\wedge \tilde e^b\wedge \tilde e^c+
\frac{c_2}{3}\,f_{abd}\L_{cd}\,\tilde e^a\wedge \tilde e^b\wedge \tilde e^c\right)\,,\\
&\widetilde\omega_{ab|c}^-=(1-\l^2)\,c_2\L_{cd}\,f_{dab}\,,\\
&\mathrm{d}\L_{ab}=c_1 f_{adc}\L_{db}\tilde e^c+c_2\left(f_{abc}-\L_{ae}f_{ebc}+\L_{ae}\L_{db}f_{edc}\right)\tilde e^c\,,\\
&\widetilde R^-_{ab}=-c_G\,c_2^2\L^{-1}_{ab}-\widetilde\nabla^-_b\om^{-c}{}_{ac}\ , \quad \widetilde\om^{-c}{}_{ac}=-c_2f_{abc}\L_{bc}\,,
\end{split}
\end{equation}
where we note that in this case $\L^{-1}=\L^T$ and that
\be
c_1=\frac{1+\l+\l^2}{1+\l}\ ,\qq c_2=\frac{\l}{1+\l}\ .
\ee
The above equal the corresponding expressions in \cite{Sfetsos:2013wia} and \cite{Itsios:2014lca}
(up to an appropriate rescaling of the vielbein and a rewriting of $\mathrm{d}\L_{ab}$).


\section{Computation of the RG flow equations}
\label{Tour}

In this section we derive the RG flow equations for \eqref{tdulalmorev2} and then study various properties and limits.

The one-loop $\b$-function equations for a general $\s$-model are given by \cite{honer,Friedan:1980jf,Curtright:1984dz}
\be
\label{RG.flows1}
{\mathrm{d} G_{\mu\nu}\ov \mathrm{d}t}+{\mathrm{d} B_{\mu\nu}\ov \mathrm{d}t}=R^-_{\mu\nu}+\nabla^+_\nu\xi_\mu\ ,
\ee
where the second term corresponds to diffeomorphisms along $\xi^\m$.
Passing to the tangent space indices with the frame $\tilde e^a= \tilde e^a_\m \mathrm{d}X^\m$ and using the definitions
\be
{\mathrm{d} G_{\mu\nu}\ov \mathrm{d}t} = \tilde  \b^G_{ab} \tilde e^a_\m \tilde  e^b_\n\ ,\qq  {\mathrm{d} B_{\mu\nu}\ov \mathrm{d}t} = \tilde \b^B_{ab} \tilde  e^a_\m \tilde e^b_\n\ ,
\ee
we have that
\be
\tilde \b^G_{ab}+\tilde \b^B_{ab}=\widetilde R^-_{ab}+ \widetilde \nabla^-_b \tilde \xi_a\ ,
\label{RG.flows}
\ee
Associate to the matrix $\tilde g_{ab}$ we also define a two-form $\tilde b_{ab}$ from
$B_{\m\n} = \tilde b_{ab} \tilde  e^a_\m \tilde e^b_\n$.

\no
We next compute the left hand side of \eqn{RG.flows}. In this computation we reinsert the parameter $k$
into the definitions of $G_{\m\n}$ and $B_{\m\n}$.
Since the WZW model term in \eqn{tdulalmorev2} does not depend on the matrix $\l$ we immediately obtain that
\be
\begin{split}
&  { \mathrm{d} G_{\m\n}\ov  \mathrm{d}t} + { \mathrm{d} B_{\m\n}\ov  \mathrm{d}t}
=   2 k { \mathrm{d}\ov  \mathrm{d}t}\left(R^a_\m (\l^{-1}-D^T)^{-1}_{ab} L^b_\n\right)
\\
&\phantom{xxxxxxxxxxx}  =  2 k R^a_\m \left[(1-\l\ D^T)^{-1}\frac{\mathrm{d}\l}{\mathrm{d}t} (1-\l D^T)^{-1}\right]_{ab} L^b_\n\ ,
\\
&
\phantom{xxxxxxxxxxx}  =  2k\,e^a_\m \left(\L^T\,{\mathrm{d}\l\ov  \mathrm{d}t} \right)_{ab} e^b_\n=
2k\,\tilde e^a_\m \left({\mathrm{d}\l\ov  \mathrm{d}t}\L^{-1} \right)_{ab}\tilde e^b_\n\ ,
\end{split}
\ee
where in the last steps we used \eqref{frames} and \eqref{Lambda}.
For completeness we note that if we had not assumed that $k$ is fixed, we would have obtained the expression
\be
\tilde\b^G_{ab}+\tilde\b^B_{ab}=2k\left({\mathrm{d}\l\ov  \mathrm{d}t}\L^{-1}\right)_{ab}+\frac{1}{k}\frac{\mathrm{d}k}{\mathrm{d}t}\left(\tilde g+\tilde b\right)_{ab}\,.
\ee
Using the latter, \eqref{Ricci.general} and \eqref{RG.flows} we conclude that
\be
\label{general.RG.final}
{\mathrm{d}\l_{ab}\ov \mathrm{d}t}=
-\frac{1}{2k}\left(\l_{\ell i}f_{a\ell p}-\l_{aq}\l_{p\ell}f_{qi\ell}\right)\left(\l_{ce}f_{bme}-\l_{nb}\l_{dm}f_{ndc}\right)g^{im}\tilde g^{pc}\ ,
\ee
where $\tilde\xi_a=\widetilde\om^{-c}{}_{a|c}$ and that $k$ does not flow.
Thus the topological nature of its quantization, due to the WZW limit (achieved when $\l\to 0$) \cite{Witten:1983ar},
persists at one-loop.

\no
We can write the above system in terms of matrices $\cN_a(\l)$ with elements
\be
(\cN_a(\l))_p{}^m = \left(\l_{ac}\l_{pd}f_{cdi} - f_{apc} \l_{ci}\right) g^{im} =:{\cN}(\l)_{ap}{}^m \ .
\ee
Then\footnote{In this rewriting of $\b$-equations, it becomes apparent that $\cN(\l)_{ab}{}^c$ play the r\^ole of deformed, by the matrix $\l$,
structure constants.}
\be
\boxed{
{\mathrm{d}\l_{ab} \ov \mathrm{d}t} = {1\ov 2k} {\rm Tr}\left(\cN_a(\l) \cN_b(\l^T)\right) = {1\ov 2k} \cN(\l)_{ap}{}^m \cN(\l^T)_{bm}{}^p
\label{dgdgg3}
}\ .
\ee
Comparing with \eqn{dhdfl} we find that the matrix $f(\l)$ has elements
\be
f_{ab}(\l) = -\ha {\rm Tr}\left(\cN_a(\l) \cN_b(\l^T)\right)\ .
\ee

\subsection{Properties of the RG flow equations}

There are several properties of the RG flow equations \eqn{dgdgg3} which we list below:

\begin{enumerate}
\item
The system \eqn{dgdgg3} satisfies the condition \eqn{symmetry.beta} due to the transformation
\be
{\cN}(\l)_{ap}{}^m \to \l^{-1}_{ac} \l^{-1}_{pd}\l_{mn}\, \cN(\l)_{cd}{}^n \ .
\ee
\item
In addition it respects the symmetry \eqref{gg0g} due to the transformation
\be
{\cN}(\l)_{ap}{}^m \to (D_0)_{ca} (D_0)_{qp}(D_0)_{nm}\, \cN(\l)_{cq}{}^n\ .
\ee
\item
It holds its form under $\l\to\l^T$.
\item
A symmetric coupling constant matrix $\l$ remains symmetric under the RG flow. This is readily seen from \eqn{dgdgg3} if we
use that $\l^T=\l$. Hence it is consistent to restrict to symmetric couplings as in \cite{Gerganov:2000mt} which
we examine closer in sections \ref{Applications} and \ref{Comparison} below.

\item
A purely antisymmetric coupling matrix is not consistent with the RG flow equations. Hence, if at some energy scale the
matrix $\l_{ab}$ is antisymmetric this will not persist along the flow.

\item
In the special case of $\l_{ab}=\l\,\d_{ab}$ or $\l_{ab}=\l\,(D_0)_{ab}$ (for constant group elements), we find that
\be
{\mathrm{d}\l\ov \mathrm{d}t}=-\frac{c_G\,\l^2}{2k(1+\l)^2}\ ,
\label{dlround}
\ee
where $c_G$ is the quadratic Casimir in the adjoint representation, defined from the relation $f_{acd}f_{bcd}= c_G \d_{ab}$.

\item
The system \eqn{dgdgg3} is in agreement with general CFT considerations
\be
{\mathrm{d}\l_{ab}\ov \mathrm{d}t} = -{1\ov 2k}\, f_{ace} f_{bdf} \l_{cd}\l_{ef} + {\cal O}(\l^3)\ ,
\label{dkhfk}
\ee
where a CFT is perturbed with operators of mass dimension equal to two as in our case.

\item
Finally, we note that \eqn{dgdgg3} encodes the RG flow equations for the non-Abelian T-dual of the PCM given in \eqn{nobag}.
We provide some details below.

\end{enumerate}

\subsection{RG flows in the general PCM and its non-Abelian T-dual}

In the limit \eqn{laborio} the system \eqn{dgdgg3} becomes
\be
\label{principal.general}
 {\mathrm{d} E_{ab}\ov \mathrm{d}t} ={1\ov 8} G^{pc} G^{mi} (E_{pl} f_{ail}
+ E_{aq} f_{qip} - E_{li} f_{alp})(E_{nb} f_{nmc} + E_{dm} f_{bdc} - E_{ce} f_{bme})\ ,
\ee
where $\displaystyle G_{ab}= \ha (E_{ab}+ E_{ba})$ and $G^{ab}=G^{-1}_{ab}$.
These are the RG flow equations corresponding to the limit
action \eqn{nobag}. Since this action is equivalent to the PCM action by a classical canonical
transformation\footnote{
This was first shown for the PCM with $E_{ab}=\d_{ab}$: For the case of $SU(2)$ in \cite{Curtright:1994be}
and for a general group
in \cite{Lozano:1995jx}. For general coupling matrix $E_{ab}$ the canonical equivalence
was established in \cite{Sfetsos:1996pm}.}
we expect that the physical information contained
in the $\b$-function equations will be preserved. That implies that the $\b$-function equations for the
general PCM should be given by \eqn{principal.general} as well.
This is already proven due to the equivalence of RG flow system of equations in
Poisson--Lie T-duality related $\s$-models in \cite{Sfetsos:2009dj}. In this context non-Abelian T-duality is a particular case.
Nevertheless, we present for completeness an independent proof that \eqn{principal.general} also follow
from the $\b$-function equations of the PCM models
action \eqn{spcm}. For this action we can derive the metric and the two-form
\be
\mathrm{d}s^2=G_{ab}\,L^aL^b\,,\qq B=\frac12\,B_{ab}\,L^a\wedge L^b\,,
\ee
with $\displaystyle G= \ha (E+ E^T)$ and $\displaystyle B=\ha (E-E^T)$
and  where we have omitted a factor of $\displaystyle {1\ov \pi}$.
It can be easily shown that the generalized spin-connections are
\be
\begin{split}
& \om^+_{ab|c}=\ha\left(E_{da}f_{dbc}-E_{cd}f_{dab}+E_{db}f_{dca}\right)\ ,\\
&
\om^-_{ab|c}=\ha\left(E_{ad}f_{dbc}-E_{dc}f_{dab}+E_{bd}f_{dca}\right)\ .
\end{split}
\ee
Using \eqref{Ricci.identity} and the latter, we can easily find that
\be
R_{ab}^-={1\ov 4} G^{pc} G^{mi} (E_{pl} f_{ail} + E_{aq} f_{qip}
- E_{li} f_{alp})(E_{nb} f_{nmc} + E_{dm} f_{bdc} - E_{ce} f_{bme})\,,
\ee
with no appearance of diffeomorphisms.
Plugging the latter into the RG flow equations
\be
{1\ov\pi}\left({ \mathrm{d} G_{\m\n}\ov  \mathrm{d}t} + { \mathrm{d} B_{\m\n}\ov  \mathrm{d}t}\right)
={1\ov 2\pi} R_{\mu\nu}^- \quad \Longrightarrow\quad
\frac{\mathrm{d}E_{ab}}{\mathrm{d}t}=\frac{1}{2}\,R_{ab}^-\,,
\ee
we readily find \eqref{principal.general}. Thus the RG flow equations of the
general PCM are the same with its non-Abelian T-dual
as stated above.

\section{Applications}
\label{Applications}

In this section we focus on cases of particular interest which involve a truncation of the form of
the matrix $\l$. This has to be done with care since setting arbitrarily entries
of the matrix $\l$ to zero will not be preserved by the flow \eqn{dgdgg3}.

\subsection{The $SU(2)$ case: Lagrange and Darboux--Halphen systems}

Consider the simplest case with $G=SU(2)$ and
\be
\l =\diag(\l_1,\l_2,\l_3)\ .
\label{llsld}
\ee
Using for representation matrices $t_a={\s_a\ov \sqrt{2}}$, where $\s_a$, $a=1,2,3$ are the Pauli matrices
leads, due to our normalization conventions, for the structure constants to
$f_{abc} =\sqrt{2}\varepsilon_{abc}$. Then from \eqn{dgdgg3} we find the system of differential equations
\be
\label{trixial2}
\frac{\mathrm{d}\l_1}{\mathrm{d}t}=-\frac2k\,\frac{(\l_2-\l_1\l_3)(\l_3-\l_1\l_2)}{(1-\l_2^2)(1-\l_3^2)}\
\ee
and cyclic in $1,2,3$. It turns out that $\displaystyle {\mathrm{d}\l_{ij}\ov \mathrm{d}t}=0$, $\forall i\neq j$,
so that the restriction \eqn{llsld} to a diagonal matrix is a consistent one.
Note the symmetry of the system under the transformation
\be
k\to -k\ ,\qq \l_i \to {1\ov \l_i} \ ,\quad i=1,2,3\ ,
\ee
which follows from \eqn{symmetry.sigma}.

\no
For $\l_i\ll 1$ this system behaves as
\be
\label{Lagrange}
{\mathrm{d}  \l_1\ov \mathrm{d} t} = - {2\ov k} \l_2 \l_3 + {\cal O}(\l^3)\
\ee
and cyclic in $1,2,3$, which is the Lagrange system.

\no
In the opposite limit, when $\l_a\to 1$, let
\be
\l_a = 1 - {x_a\ov k} + {\cal O}\Big({ 1\ov k^2}\Big)\ ,\qq a=1,2,3\ .
\label{llh1}
\ee
Then in the limit $k\to \infty$ we obtain
\be
\label{DH}
{\mathrm{d} x_1\ov \mathrm{d} t}=1 + {1\ov 2 x_2 x_3} (x_1^2 -x_2^2 - x_3^2)\
\ee
and cyclic in $1,2,3$, which is the Darboux--Halphen system. This system also follows from \eqn{principal.general} with
$E_{ab}=\diag(x_1,x_2,x_3)$.
It first appeared in RG flows for the PCM for $SU(2)$ with the above diagonal matrix $E_{ab}$ in \cite{Cvetic:2001zx}.
The Lagrange and the Darboux--Halphen systems arose before in general relativity by imposing the
self-dual condition on the Bianchi IX with $SU(2)$ isometry four-dimensional Euclidean metrics \cite{Gibbons:1979xn}.

In conclusion the system \eqn{trixial2} interpolates between the Lagrange and the Darboux--Halphen systems.
These have a Lax pair formulation, i.e. \cite{Takhtajan:1992qb}.
It is very interesting to investigate if this is the case for the interpolating system \eqn{trixial2} as well.

\subsection{The two coupling case using a symmetric coset}

Let's split the group index into a part corresponding to a subgroup $H$ of $G$ and the
rest belongings to the coset $G/H$. We will keep denoting by Latin letters the subgroup indices and by
Greek letters the coset indices.
Consider the case in which the matrix $\l$ has elements
\be
\l_{ab} = \l_H \d_{ab}\ ,\qq \l_{\a\b} = \l_{G/H} \d_{\a\b}\ .
\ee
It turns out that the above restriction is consistent with the system \eqn{dgdgg3} only for symmetric coset spaces $G/H$,
for which the structure constants $f_{\a\b\g}=0$.
Using that for symmetric spaces
\be
f_{abc} f_{abd} = c_{H}\d_{cd}\ ,\quad f_{\a\b c} f_{\a\b d} = (c_{G}-c_{H})\d_{cd}\ ,
\quad f_{c\b\g} f_{c\b\d} = {c_{G}\ov 2} \d_{\g\d}\ ,
\label{casim}
\ee
we find the system of equations
\be
\begin{split}
\label{RG.sub.coset2}
&\frac{\mathrm{d} \l_H}{\mathrm{d} t}=-\frac{c_G\l_{G/H}^2(1-\l_H^2)^2+c_H(\l_H^2-\l_{G/H}^2)(1-\l_H^2\l_{G/H}^2)}{2k(1+\l_H)^2(1-\l_{G/H}^2)^2} \,,\\
&\frac{\mathrm{d} \l_{G/H}}{\mathrm{d} t}=-\frac{c_G\l_{G/H}(\l_H-\l_{G/H}^2)}{2k(1+\l_H)(1-\l_{G/H}^2)} \,.
\end{split}
\ee
In the limit of $\l_{G/H}=0$  they consistently truncate to the RG flow equation \eqn{dlround} with $(\l,c_G)$ replaced by $(\l_H,c_H)$.\footnote{\label{general.coset}
In fact this is the case for general cosets and for coupling matrices of the block
diagonal form $\l=\l_H\oplus{\mathbb O}_{G/H}$, where $\l_H$ is a general $\dim H$ square matrix.
To prove this we observe that the non-vanishing components of the matrices ${\cN}_m$ are
\begin{equation*}
(\cN_a)_b{}^c =\left(\l_{ad}\l_{be}f_{def} - f_{abe} \l_{ef}\right) g^{fc}\ ,\qq (\cN_\a)_\b{}^c =- f_{\a\b d} \l_{de}\, g^{ec}\ .
\end{equation*}
Then clearly the RG flow equations \eqn{dgdgg3} become that for a subgroup $H\in G$ with $\l$ replaced by $\l_H$.
Consistent to these is the fact that the action \eqn{tdulalmorev2}
retains its form but with $\l$ replaced by $\l_H$ and with the indices $a,b$ taking values in $H$. The interpretation of this new more
general  than \eqn{tdulalmorev2} action is of the all-loop action of the anisotropic non-Abelian Thirring model for a group $G$
where the perturbation is by a general current bilinears  in $H\in G$.}
\no
For small couplings these read
\be
\begin{split}
\label{RG.sub.cosetlimit}
& \frac{\mathrm{d} \l_H}{\mathrm{d} t}=
- {1\ov 2 k}\left(c_H \l_H^2 + (c_G-c_H) \l_{G/H}^2\right)+ {\cal O}(\l^3)\ ,\\
& \frac{\mathrm{d} \l_{G/H}}{\mathrm{d} t}=
- {c_G\ov 2 k} \l_H \l_{G/H} + {\cal O}(\l^3) \ ,
\end{split}
\ee
which is in agreement with general CFT expectations, i.e. with \eqn{dkhfk}.
Note that for groups for which their rank can be taken arbitrarily large
the exact expression for the running of couplings
can be simply obtained from the perturbative result. This can be seen by noting that
when $c_G\to \infty$ and $c_H\to \infty$
so that their ratio remains finite we may define $x_H= c_G \l_H$ and $x_{G/H}= c_G \l_{G/H}$.
Then the running of the $x$'s coincides
with that we would have obtained from the leading terms in \eqn{RG.sub.cosetlimit}.

\subsubsection{Flows from coset CFTs}

Recall that in \cite{Sfetsos:2013wia} a second class of $\s$-models was constructed interpolating between coset CFTs realized by
gauged WZW models and the non-Abelian T-dual of coset PCM models.
This construction is based in a limiting
procedure in which the part of the coupling matrix $E$ (taken to have a block diagonal structure in the subgroup $H$ and the
coset $G/H$ spaces) in \eqn{skE} with subgroup indices is taken to be zero.
Then the action \eqn{tdulalmorev2} depends actually not on all of the $\dim(G)$ variables $X^\m$ but on $\dim(G/H)$ variables.
For more details the interested reader is refereed to \cite{Sfetsos:2013wia}.
Returning to our case note that in this decoupling limit of the subgroup the parameter $\l_H=1$. Then we find from \eqn{RG.sub.coset2} that
\be
\label{RG.sub.coset.1}
\frac{\mathrm{d} \l_{G/H}}{\mathrm{d} t}=-\frac{c_G\l_{G/H}}{4k} \ ,
\ee
which again is in agreement with general CFT expectations.
It is interesting that the all-loop result is identical to the one-loop in $\l_{G/H}$ perturbative
CFT result, at leading-order in the $1/k$ expansion.
The result in \eqn{RG.sub.coset.1} is essentially the same as that obtained in \cite{Itsios:2014lca}
for the simplest case with $G=SU(2)$ and $H=U(1)$. In addition the
RG flow will be the same no matter what the subgroup
$H$ is, as long as the coset $G/H$ is a symmetric space.
We note that $\l_H=1$ is a fixed point, only if the subgroup is Abelian,
see also \cite{Sfetsos:2009dj}.\footnote{
It turns out that $\l_H=1$ corresponds to a fixed point of the group $G$ since $\l_{G/H}=1$
is also enforced.
To prove this, we use \eqref{RG.sub.coset2} for $\l_H\to1^-$ and we find for Abelian subgroups that
\begin{equation*}
{\mathrm{d}\l_{G/H}\ov \mathrm{d}\l_H}\simeq {(1-\l_{G/H}^2)^2\ov 2 \l_{G/H}(1-\l_H)^2}
\quad \Longrightarrow\quad
\l_{G/H}\simeq1-\frac{1-\l_H}{2}+{\cal O}(1-\l_H)^2\,.
\end{equation*}
Hence, the $\s$-model flows towards the IR to the non-Abelian T-dual of the PCM for the group $G$.
}

\section{Comparison with literature}
\label{Comparison}

The purpose of this section is to compare results following from our general formula for the $\b$-functions \eqn{dgdgg3}
with existing ones in the literature.

The authors in \cite{Gerganov:2000mt,LeClair:2000ya}, considered the anisotropic Thirring model action
given by the first two terms in \eqn{thorio} with symmetric coupling matrix $\l$ and computed the corresponding $\b$-functions using
current algebra CFT techniques. The general formula they obtained is not apparently the same as \eqn{dgdgg3} and in fact it looks more complicated.
Given our completely different approach it is important to make a comparison. First we briefly review
the results of \cite{Gerganov:2000mt}. One considers a perturbation of the form
\be
S_{\rm pert} = \int \sum_A h_A {\cal O}^A\ ,\qq {\cal O}^A  = \sum_{a,b=1}^{\dim G} d^A_{ab} J_+^a J_-^b\ ,
\label{perth}
\ee
where $d^A_{ab}$ are pure numbers and define the perturbation. In this work the $d^A_{ab}$'s were
taken to be symmetric in the lower indices $a,b=1,2,\dots,\dim(G)$.
The upper index $A$ takes as many values as the number of independent coupling constants $h_A$
(denoted by $g_A$ in \cite{Gerganov:2000mt}). Hence,
comparing with our notation we have the identification $\l_{ab} = h_A d^A_{ab}$.
Due to that $d^A_{ab}=d^A_{ba}$ the comparison with our results can only be made for
symmetric matrices $\l$.

\no
One assumes that the operators ${\cal O}^A$ form a closed set. Then, from the double pole in the
operator product expansion
\be
{\cal O}^A(z,\bar z) {\cal O}^B(0,0) = {1\ov |z|^2} {\cal C}^{AB}{}_C {\cal O}^C(0,0) + \cdots \ ,
\ee
one extracts the structure constants ${\cal C}^{AB}{}_C$. The following three conditions ensure closeness of this
algebra and renormalizability at all orders
\be
d^A_{ab} d^B_{cd} f_{ace} f_{bdf} ={\cal C}^{AB}{}_C d^C_{ef}\ ,\qq
d_{ac}^A d^B_{bc} = {\cal D}^{AB}{}_C d^C_{ab}\ ,\qq d^A_{cd} f_{a e c} f_{e b d} = {\cal R}^A{}_B d^B_{ab}\ .
\ee
These relations define the set of coefficients ${\cal D}^{AB}{}_C$ and ${\cal R}^A{}_B$. Note also the consistency relations
\be
{\cal C}^{AB}{}_C = {\cal C}^{BA}{}_C\ ,\qq {\cal D}^{AB}{}_C = {\cal D}^{BA}{}_C\ ,\qq {\cal D}^{AC}{}_D {\cal D}^{DB}{}_E = {\cal D}^{AB}{}_D {\cal D}^{DC}{}_E\ .
\ee
Finally one defines the vector and the matrix
\be
{\cal C}_A(x,y)={\cal C}^{BC}{}_A x_B y_C\ ,\qq {\cal D}^A{}_B = {\cal D}^{AC}{}_B h_C\ ,
\label{cvv}
\ee
for any two vectors $x$ and $y$, as well as
\be
\tilde h_A = h_B ((\mathbb{I}-{\cal D}^2)^{-1})^B{}_A  \ .
\ee
Then the $\beta$-function equations are given by \cite{Gerganov:2000mt,LeClair:2000ya}
\be
{\mathrm{d}h_A\ov \mathrm{d}t} = {1\ov k} \left[\ha {\cal C}_B(\tilde h,\tilde h) (\mathbb{I} + {\cal D}^2)^B{}_A
- {\cal C}_B(\tilde h{\cal D},\tilde h{\cal D}) {\cal D}^B{}_A  + \tilde h_B ({\cal D}{\cal R}{\cal D})^B{}_A \right]\ ,
\label{dhadt}
\ee
where $(\tilde h{\cal D})_A = \tilde h_B {\cal D}^B{}_A$.
Finally we note that it is consistent to truncate to a subgroup $H\in G$, by considering $d^A_{ab}$'s with lower indices only in the
Lie algebra of $H$. This is congruous with the discussion in footnote \ref{general.coset}.

In the following we concentrate on two non-trivial examples, namely the $SU(2)$ and the symmetric
coset $G/H$ cases, where we will use \eqn{dhadt} to compute explicitly
the RG flow equations for the couplings. We will find perfect agreement with \eqn{trixial2} and \eqn{RG.sub.coset2}
we have found using our general formula \eqn{dgdgg3}. Based on that we believe that the system \eqn{dhadt} becomes identical to that in
\eqn{dgdgg3} for the case of a symmetric, but otherwise general, coupling matrix $\l$. The proof should be
a nice mathematical exercise.

\subsection{The $SU(2)$ case}

For the case of $SU(2)$ consider turning on just three couplings, $h_A$, $A=1,2,3$ a case that has been
considered in \cite{LeClair:2004ps}.
Referring to \eqn{perth} and choosing as the non-vanishing $d^A_{ab}$ those with
\be
d^1_{11}=d^2_{22}=d^3_{33}=1\ ,
\ee
implies, by comparing with \eqn{llsld}, that $\l_1=h_1 $, etc.
Then the non-vanishing ${\cal C}^{AB}{}_C$'s, ${\cal D}^{AB}{}_C$'s and ${\cal R}^A{}_B$'s are
\begin{equation}
\begin{split}
& {\cal C}^{12}{}_3 = {\cal C}^{21}{}_3= - 2 \ ,\qq {\cal D}^{11}{}_1 =1 \ ,\qq {\cal R}^1{}_2 = {\cal R}^2{}_1 =2 \ ,\\
& {\cal C}_1(x,y)=-2( x_2 y_3 + x_3 y_2) \ ,\qq {\cal D}^1{}_1 = \l_1
\end{split}
\end{equation}
and cyclic permutations in $1,2,3$. Then
plugging these into \eqn{dhadt} we find \eqn{trixial2}, a result consistent with that in \cite{LeClair:2004ps}
(after relabeling and rescaling).

\subsection{The two coupling case using a symmetric coset}

\no
Let us concentrate on the case of symmetric coset spaces for which $f_{\a\b\g} = 0$.
We turn on two couplings,
one for the subgroup $h_H$ and one for the coset $h_{G/H}$. As before let Latin indices denote the subgroup and Greek ones the coset.
Hence, we have that
\be
d^1_{ab} = \d_{ab} \ ,\qq d^2_{\a\b}= \d_{\a\b}\ ,
\ee
where 1 and 2 refer to the subgroup and coset, respectively. Then the non-vanishing ${\cal C}^{AB}{}_C$'s and ${\cal D}^{AB}{}_C$'s are
\be
\begin{split}
& {\cal C}^{11}{}_1 = c_{H}\ ,\qq {\cal C}^{12}{}_2= {\cal C}^{21}{}_2= {c_{G}\ov 2}\ , \qq {\cal C}^{22}{}_1 = c_{G}-c_{H}\ ,\\
& {\cal D}^{11}{}_1 = {\cal D}^{22}{}_2 = 1\ ,
\end{split}
\ee
where $c_G,c_H$ are the quadratic Casimir in the adjoint representation of $G$ and $H$, respectively.
Also
\be
{\cal R}^1{}_1 = -c_{H} \ ,\qq {\cal R}^1{}_2= {\cal R}^2{}_2 = -{c_{G}\ov 2}\ ,\qq {\cal R}^2{}_1 = -c_{G}+c_{H}\ .
\ee
In deriving the above we used \eqn{casim}.
Plugging these into \eqn{dhadt} we find precisely \eqn{RG.sub.coset2} with the identification $\l_H= h_H$ and $\l_{G/H}=h_{G/H}$.

\subsubsection*{Non-diagonal case}

Finally, the reader might also require a further example, involving  a non-diagonal (symmetric) coupling matrix, so to strengthen the declared equivalence
between \eqn{dgdgg3} and \eqn{dhadt}. We did so for a consistent truncation in the $SU(2)$ case, i.e. $\l_{12}=\l_{21}$ and $\l_{33}$ and the
results followed from these two expressions are in perfect agreement.

\section{Conclusion and outlook}
\label{conclusion}

The main result of the present paper is the proof of the one-loop renormalizability and the
computation of the RG flow equations for the coupling matrix $\l_{ab}$ of
the action \eqn{tdulalmorev2}.
This computation was achieved using the standard expression for the one-loop renormalizability
of two-dimensional $\s$-models of \cite{honer,Friedan:1980jf,Curtright:1984dz} and the
explicit expression for the $\beta$-function equations is given in \eqn{dgdgg3}.
We used this result to further claim that the action \eqn{tdulalmorev2} is the effective
action encoding all loop effects in the coupling matrix $\l_{ab}$ of the fully anisotropic non-Abelian Thirring model action defined by the
first two terms in \eqn{thorio}.
The basic support for the above claimed equivalence is the fact that in some highly non-trivial cases our RG flow equations are the same as
the ones we have found using a general formula for the running of couplings of the anisotropic symmetric non-Abelian Thirring model action
given by \eqn{dhadt}.
This formula was obtained in \cite{Gerganov:2000mt} using current-algebra techniques and is an all order in the couplings result (but leading in $1/k$).
Even though we have not proven the equivalence of \eqn{dgdgg3} and \eqn{dhadt} in general we believe this to be the case for the case
(of course when $\l_{ab}$ is symmetric). Our result \eqn{dgdgg3} has the advantage of being more general since it includes cases of
non-symmetric $\l_{ab}$ and in addition it has a remarkably much simpler form.
It will be interesting to investigate the RG flow equations \eqn{dgdgg3} for specific low dimensional groups.

\no
We have found a particularly interesting result for the anisotropic $SU(2)$ case with diagonal coupling matrix $\l_{ab}$.
The RG flow equations interpolate between the Lagrange and Darboux--Halphen integrable systems of
differential equations. It will be interesting if the integrability property of the system is maintained in general beyond the
these limit cases.

Finally we note the existence of the all-loop RG flow equations for the anisotropic bosonized non-Abelian Thirring model
with different left and right levels of the current algebra \cite{LeClair:2001yp}.
In particular, using  Eq.(4.6) of this work we have computed the $\b$-function for left and right currents with different levels $k_L$ and
$k_R$. The result is
\begin{equation}
{
\frac{\mathrm{d}\l_1}{\mathrm{d}t}=-\frac{2(1+k_Lk_R\l_1^2)\l_2\l_3-(k_L+k_R)\l_1(\l_2^2+\l_3^2)}{(1-k_Lk_R\l_2^2)(1-k_Lk_R\l_3^2)}\
}
\end{equation}
and cyclic in $1,2$ and $3$.
This is the analogue of \eqn{trixial2} to which reduces for the special case with $k_L=k_R=k$
and $\l_i\mapsto \l_i/k$.
We have also computed the analogue of \eqn{RG.sub.coset2} for the case of symmetric cosets $G/H$ with different levels for
the left and right current algebras but will not present here the result.
Such models are not captured
by our all-loop action Eq.\eqn{tdulalmorev2}, since the WZW action provides a left and a right current algebra
with equal levels. It is will be interesting to realize the above RG flow systems of equations with specific $\s$-models.

As mentioned, for the general $\s$-model action \eqn{tdulalmorev2} only the case with $\l_{ab}=\l\d_{ab}$
corresponding to \eqn{tdulalmorev2orio}
has been proven to be integrable.
It is likely that other choices for $\l_{ab}$ may correspond to integrable models as well.
A necessary condition  for a model to be integrable is that the non-Abelian action \eqn{nobag} to which it tends in the
limit \eqn{laborio} is integrable. Since non-Abelian T-duality preserves integrability (see appendix D of \cite{Sfetsos:2013wia})
this is equivalent to looking at cases with proven integrability in PCM which have been worked out in the literature.
Such a case is the anisotropic with diagonal $\l_{ab}$ coupling matrix $SU(2)$ PCM
for which integrability was shown in \cite{Cherednik:1981df,hlavaty,Mohammedi:2008vd}.
It will be highly non-trivial if this model is integrable.

\section*{Acknowledgements}
The research of K.\,Sfetsos is implemented
under the \textsl{ARISTEIA} action (D.654, GGET) of the \textsl{operational
programme education and lifelong learning} and is co-funded by the
European Social Fund (ESF) and National Resources (2007-2013). The work of K.
Siampos has been supported by  \textsl{Actions de recherche
concert\'ees (ARC)} de la \textsl{Direction g\'en\'erale de
l'Enseignement non obligatoire et de la Recherche scientifique
Direction de la Recherche scientifique Communaut\'e fran\c{c}aise de
Belgique} (AUWB-2010-10/15-UMONS-1), and by IISN-Belgium (convention 4.4511.06). The authors
would like to thank each others home institutions for hospitality, where part of this work was developed.

\appendix


\end{document}